# A Survey of Maturity Models from Nolon to DevOps and Their Applications in Process Improvement

James J. Cusick, PMP
IEEE Computer Society Member
*j.cusick@computer.org*
New York, NY

*Abstract*—This paper traces the history of Maturity Models and their impact on Process Improvement from the early work of Shewhart to their current usage with DevOps. The history of modern process improvement can be traced at least to Shewhart. From his foundational process contributions and those of other innovators a variety of methods and tools to aid in process quality advancement were developed. This paper begins by reviewing those early steps and then focuses on the emergence of Maturity Models in the 1970s with Nolan's initial approach. The broad adoption of Maturity Models that followed through the success of the CMM and then the CMMI approaches is detailed. This then leads to a general survey of additional models developed for such areas as IT Service Management, ITIL, Project Management, Agile Development, DevOps, CERT, and MDM among others. Finally, this paper discusses the application of these models in the support of process improvement and their limitations. Readers of this paper can expect to gain an appreciation for the origins of these models and surrounding methods as well as an ability to conduct comparative analysis of such models to aid in their selection and application.

## I. Introduction

Recently while developing a strategy to conduct an organizational process baseline it became apparent that it was necessary to define several terms around what was commonly understood by management to be a "CMMI assessment". CMMI (Capability Maturity Model Integrated) [Chrissis, 2003] is well understood in Software Engineering circles for which it was originally developed. However, it has also been appropriated as a term to describe the process or organizational maturity of many kinds of IT or business functions. It is the intent of this paper to describe the origins of the broader term "management maturity model" (MMM), its various forms, uses, and limits. This paper will also provide a survey of several kinds of alternative maturity models including those available for ITSM/ITIL, Agile, DevOps, Project Management, and more. This understanding of the history, evolution, and usage of maturity models can help projects and process engineers in evaluating which model to apply in a given environment and for a given purpose. Finally, this review can guide readers on how to leverage these models in conducting deep process improvement activities over time leading to increasing capabilities and quality for customers.

## II. Process Improvement Roots

There is a long history around process improvement which eventual leads to the emergence of maturity models. Early approaches stem from the Scientific Method (Taylor, 1911) and empirical approaches to natural history (Galileo, 1638; Bacon, 1696). A foundational contribution in the modern era came from Dr. Walter A. Shewhart of Bell Labs in 1924. Shewhart introduced the concept of a Statistical Process Control Chart (SPC) which is used to demonstrate quality improvement over time as it relates to changes in input, process, or materials. This work was later documented with supporting statistical methods in his 1931 work "Economic Control of Quality of Manufactured Product" [Shewhart, 1925]. This introduced many core concepts around improving processes and developing process maturity.

A key collaborator of Shewhart's was Joseph Juran who, while at the Hawthorne Works of AT&T's Western Electric, worked with Shewhart and the Bell Labs teams on using SPCs and other methods of quality engineering. Juran later taught these quality methods in post-war Japan which led to the well-known improvement cycle which the Japanese called the "Plan-Do-Check-Act" approach but is also known as the "Shewhart Cycle" (see Figure 1) [Juran, 1951]. Both Juran from 1951 and later Edward Deming (who had also worked with Shewhart) transferred these core improvement and quality concepts broadly to Japanese researchers, managers, and engineers [Deming, 1982].

Subsequently, Total Quality Management (TQM) emerged from several origins but with the consolidated authorship of Ishikawa in Japan [Ishikawa, 1968]. Within TQM top management typically initiates change with the participation of all employees. Change and improvement in TQM includes adoption of the TQM philosophy; implementation of SPC; and running of "Plan-Do-Check-Act" cycles tying many of the ideas of Shewhart, Juran, and Deming together [Ishikawa, 1985]. The origin of these methods is seen in Figure 1 from Moen [2010]. Later, TQM then spawned several other quality approaches including Lean Manufacturing and Six Sigma. Today even "modern" development approaches trace some of their ideas to this legacy and have directly embraced methods like Lean. This is especially true with Agile Development. Many of these methods underlie the approaches of the maturity models reviewed below. Importantly, the core concepts of incremental improvement and measurement became fundamental to engineering and development within industry. This ongoing cyclical process then led to the eventual structuring of categories of capabilities in ladders of maturity or "maturity models".

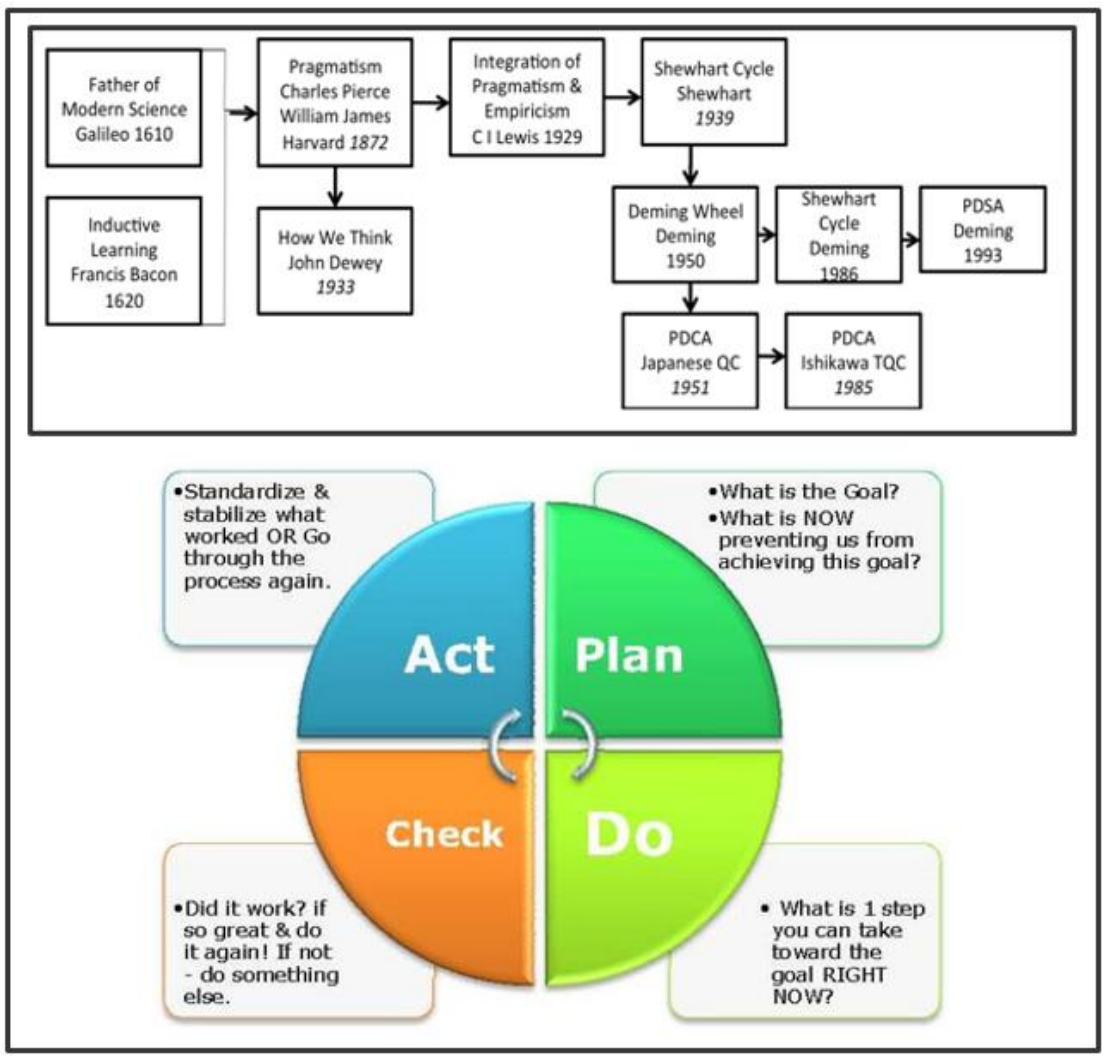

*Figure 1 – The History and Structure of the PDCA Cycle*

III. MANAGEMENT MATURITY MODELS

By the 1970s quality issues in US manufacturing and the obvious superiority of Japanese quality had people looking for answers and alternatives. Quality leaders like Deming then found an audience back home in the US consulting on the same statistical improvement methods they had taught in Japan. These quality movements eventually formed the TQM (Total Quality Management) approach. [ASQ, 2019]. Others also began looking at systemic issues for management systems especially in IT to deliver on quality. In a business context, maturity became defined as a measurement of the ability of an organization for continuous improvement in a particular discipline. Most maturity models assess qualitatively people/culture, processes/structures, and objects/technology [von Wangenheim, 2010].

*A. Nolan's Organizational Management Stages*

The first such formalized approach came from Richard Nolan in 1973. In a key article published in the Harvard Business Review in 1973 [Nolan, 1973] Nolan laid out a model for improving the operational maturity of an organization's maturity. While he never uses the term "maturity model" this can be seen as the de facto introduction of the concept. Nolan traced business situations where scope and complexity were growing much faster than the ability to control IT development and operations. His solution to this was a 6-stage evolution path from less control to the opposite where the organization had a higher level of control (See Figure 2). This model shows a remarkable similarity to current understandings of maturity models.

**Exhibit II — Optimum balance of organizational slack and control**

| Stages | Organizational slack — Computer Data | | Control — Computer Data | | Objective of control systems |
|---|---|---|---|---|---|
| Stage 1 | Low | | Low | | |
| Stage 2 | High | | Low | | Facilitate growth |
| Stage 3 | Low | Low | High | Low | Contain supply |
| Stage 4 | | High | | Low | Match supply and demand |
| Stage 5 | Low | | High | | Contain demand |
| Stage 6 | | High | | High | Balance supply and demand |

*Figure 2 – Nolan's organizational evolution model to higher IT control from 1973 – a forerunner of the management maturity model.*

*B. Crosby's Quality Grid*

Just a few years after Nolan's paper, Phil Crosby (who developed the concept of "zero defects"), published his landmark book "Quality is Free" [Crosby, 1979]. This book-marked decades of experience and study on what quality was all about in product development. A core part of the book was his "Quality Management Maturity Grid" (see Figure 3). The model abbreviated as QMMG is credited with being the precursor maturity model for the Capability Maturity Model (CMM) which was created a decade later. Like the QMMG the CMM also has five levels of maturity. This model directly built on the process improvement methods pioneered by Shewhart and expanded by Juran, Deming, and Ishikawa.

Crosby's model, aside from being highly influential and built on existing quality principles was also intuitive and thus provided powerful insights for managers into the nature and performance of their organizations. This also provided a means to map the future indicating where improvements were called for and providing much deeper direction than Norton's approach. Thus, this was the emergence of the first robust capability model as opposed to improvements focused on point in time production or operations issues. This then set the wheels in motion for the next generation maturity models appearing only a few years into the future.

*C. The Emergence of CMM and CMMI*

Following Crosby's work on his Quality Maturity Grid others began looking at this approach to organize quality improvement initially in the software domain. Essentially, the early quality work starting with Shewhart and leading to many modern methods now found an organizing principle to conduct dead reckoning (assessments) and orchestrate comprehensive quality improvements through staged process improvements in the IT domain.

| Quality Management Maturity Grid (Crosby) | Assessor: | | Department: | | |
|---|---|---|---|---|---|
| Measurement Categories | Stage 1: Uncertainty | Stage 2: Awakening | Stage 3: Enlightenment | Stage 4: Wisdom | Stage 5: Certainty |
| Management understanding and attitude | No comprehension of quality as a management tool. Tend to blame quality department for "quality problems". | Recognising that quality management may be of value but not willing to provide money or time to make it all happen. | While going through quality improvement programme learn more about quality management; becoming supportive and helpful. | Participating. Understand absolutes of quality management. Recognise their personal role in continuing emphasis. | Consider quality management as an essential part of company system. |
| Quality organisation status | Quality is hidden in manufacturing or engineering departments. Inspection probably not part of organisation. Emphasis on appraisal and sorting. | A stronger quality leader is appointed but main emphasis is still on appraisal and moving the product. Still part of manufacturing or other. | Quality department reports to top management, all appraisal is incorporated and manager has role in management of company. | Quality manager is an officer of company; effective status reporting and preventive action. Involved with customer affairs and special assignments. | Quality manager on board of directors. Prevention is main concern. Quality is a thought leader. |
| Problem handling | Problems are fought as they occur; no resolution; inadequate definition; lots of yelling and accusations. | Teams are set up to attack major problems. Long-range solutions are not solicited. | Corrective action communication established. Problems are faced openly and resolved in an orderly way. | Problems are identified early in their development. All functions are open to suggestion and improvement. | Except in the most unusual cases, problems are prevented. |
| Cost of quality as % of sales | Reported: Unknown Actual: 20% | Reported: 3% Actual: 18% | Reported: 8% Actual: 12% | Reported: 6.5% Actual: 8% | Reported: 2.5% Actual: 2.5% |
| Quality improvement actions | No organised activities. No understanding of such activities | Trying obvious "motivational" short-range efforts. | Implementation of a multi-step programme (e.g. Crosby's 14-step) with thorough understanding and establishment of each step. | Continuing the multi-step programme and starting other pro-active / preventive product quality initiatives. | Quality improvement is a normal and continued activity. |
| Summary of company quality posture | "We don't know why we have problems with quality". | "Is it absolutely necessary to always have problems with quality?" | "Through management commitment and quality improvement we are identifying and resolving our problems." | "Defect prevention is a routine part of our operation." | "We know why we do not have problems with quality." |

*Figure 3 - Crosby's Quality Management Maturity Grid – The first formal maturity model.*

The initial and most successful such model was the Capability Maturity Model (CMM). This model was first developed by Rob Radice and Watts Humphrey at IBM Federal Systems working with the US military [Radice, 1985]. This model provided for the following five point scale: 1) Traditional; 2) Awareness; 3) Knowledge; 4) Skill & wisdom; and 5) Integrated management system. Humphrey then based the CMM on Crosby's model but modified it to solve process issues in a particular order building on the IBM experience. Work on the CMM model took place mostly in 1985-1986 culminating with a release of the framework by the Software Engineering Institute when Humphrey joined the SEI in 1986. The first public version of the CMM was published in 1988 and quickly became widely adopted [Paulk, 2001]. Later versions refined the model in 1991 and in book form in 1995. A significant part of the success of CMM can be attributed to the institutional support and promotion from the SEI as well as the decision by the US Air Force to adopt the CMM model as a requirement for software suppliers [Paulk, 1994]. The progression of maturity levels of both the CMM and the later CMMI are shown comparatively in Figure 4 below.

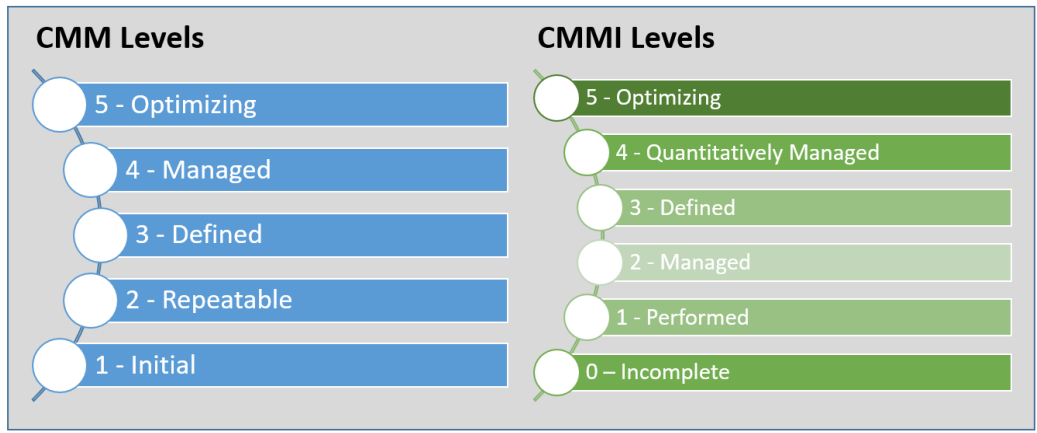

*Figure 4 – CMM and CMMI Maturity Level Labels Compared.*

Working from the experiences with CMM implementation and due to various limitations in the application of CMM the SEI developed the CMMI (Capability Model Maturity Integration) [Chrissis, 2007]. The key difference between CMM and CMMI is that CMMI allows for either a staged or a continuous approach. This gives users flexibility in applying the process maturity model. There were numerous other improvements to the model as well (see Figure 5 for the capabilities of the CMMI levels). Few people apply CMM currently. Those that do use these models tend to apply CMMI which is considered to have superseded CMM. The newer CMMI framework also began supporting other disciplines outside of Software Engineering which now include Service and Supplier Management, for example. Implicit in the CMM/CMMI framework is the use of supporting process improvement methods such as SPCs guided by PDCA cycles. Again, this brings us full circle from Shewhart to Ishikawa to Crosby & Humphrey.

| Level | Character | Description | Capabilities | Result |
|---|---|---|---|---|
| 5 Optimizing | Focus on process improvement | Continuous Process Improvement | • Organizational Innovation & Deployment<br>• Casual Analysis & Resolution | Productivity and Quality |
| 4 | Quantitatively Managed | Processes measured and controlled. | Quantitative Management | • Quantitative Process Management | |
| 3 | Defined | Processes characterized for the organization and is proactive | Process Standardization | • Organizational Process Focus<br>• Organizational Process Definition<br>• Organizational Training<br>• Integrated Supplier Management<br>• Integrated Supplier Management | |
| 2 | Managed | Processes characterized for projects and is often reactive. | Basic Project Management, many times reactive | • Requirements Management<br>• Supplier Agreement Management<br>• Measurement and Analysis | |
| 1 | Initial | Processes unpredictable, poorly controlled and reactive | Firefighting, heroic efforts | • Design<br>• Develop<br>• Integrate<br>• Test | Risk and Waste |

*Figure 5 - CMMI Reference Model (See ISACA for details on CMMI: https://cmmiinstitute.com/).*

D.  Current Generation CMMI Models

The original motivation for this paper was to clarify the meaning of the types of maturity models available that could guide operations process assessments. A commonly cited model was CMMI. Importantly, there are several types of CMMI models to choose from as mentioned. These range from the CMMI Dev, CMMI SVC (Figure 6 for example), CMMI ACQ, and CMMI DMM. Additionally, CMMI 2.0 now accounts for Agile and DevOps methodologies [CMMI, 2004]. This assumes that a CMMI framework is in fact desired as the assessment framework. Using a CMMI model brings with it significant commitment and complexity. This also can call for an assessment by a certified assessor which can be expensive on its own not counting the work which the organization must do in process preparation to meet their objectives within the model. As it turned out with the project at hand, a CMMI assessment was not actually expected. However, for convenience the assessment was labeled a "CMMI" assessment out of terminology familiarity among management. Instead, the challenge became selecting from the right MMMs that were available. This brings us to looking at some alternative current day maturity models.

| Level | Focus | Process Areas | Quality Productivity |
|---|---|---|---|
| 5 Optimizing | Continuous Process Improvement | Organizational Performance Management (OPM)<br>Causal Analysis and Resolution (CAR) | |
| 4 Quantitatively Managed | Quantitative Management | Organizational Process Performance (OPP)<br>Quantitative Work Management (QWM) | |
| 3 Defined | Process Standardization | Capacity and Availability Management (CAM) (svc)<br>Incident Resolution and Prevention (IRP) (svc)<br>Service System Transition (SST) (svc)<br>Service Continuity (SCON) (svc)<br>Service System Development (SSD) (svc, optional)<br>Strategic Service Management (STSM) (svc)<br>Organizational Process Focus (OPF)<br>Organizational Process Definition (OPD)<br>Organizational Training (OT)<br>Integrated Work Management (IPM)<br>Risk Management (RSKM)<br>Decision Analysis and Resolution (DAR) | |
| 2 Managed | Basic Project Management | Service Delivery (SD) (svc)<br>Requirements Management (REQM)<br>Work Planning (WP)<br>Work Monitoring and Control (WMC)<br>Supplier Agreement Management (SAM)<br>Measurement and Analysis (MA)<br>Process and Product Quality Assurance (PPQA)<br>Configuration Management (CM) | Risk Rework |
| 1 Initial | | | |

*Figure 6 – CMMI-SVC (Services) – An example of the expanded scope of CMMI model domains.*

## IV. ALTERNATIVE MATURITY MODELS

Following the success of the CMM and CMMI models' researchers and quality organizations quickly developed alternative maturity models to cover many areas of IT and business. The proliferation of these models has produced models covering a wide array of domains but not always with equal depth or rigor in supporting practices such as assessments and process improvement as was provided initially by CMM. We can see a straight line from Crosby's quality grid to the CMM to many of the following models:

- Performance Management CMM
- Project Management CMM
- Quality CMM
- Business Process CMM
- CERT CMM
- Enterprise IT Performance CMM
- ITIL Maturity Model
- DevOps Maturity Model
- People Capability Maturity Model
- BigData Capability Maturity Model
- Cybersecurity Capability Maturity Model
- Systems Engineering CMM
- Software Acquisition CMM
- ISO/IEC 15504
- MD3M (Master Data Model Maturity Model)
- Sustainability Capability Maturity Model

As is obvious, the straightforward nature of the maturity model concept has spawned these various implementations in several fields. Their popularity keeps many of them alive and new applications of the construct keeps being found. Using CMMI more as a label we can then consider the actual maturity models which would need to be applied to properly assess a given organization from all the available models. While there are many models to choose from, once we start reviewing these models for applicability there are a few that emerge as the most applicable to a given organization and its process environment and goals. As an example, the following models are reviewed in some detail below. Additional models can also be reviewed as needed. Looking at the set of models that need to be considered for use it would make sense to include the following for assessment purposes:

1. CMMI – (covered above).
2. ITSM/ITIL Maturity Model.
3. Agile Maturity Model.
4. DevOps Maturity Model.
5. MDM Maturity Model.

There may be some overlap between these models, so it would be recommended to reduce the number of models and to simplify any assessment approach. For some teams they may in fact want to use the CMMI-DEV or CMMI-Agile frameworks depending on what applies. However, if the team handles IT Operations the ITIL model would be better suited. Of course, the Agile model should be used for those teams following such methods as Scrum. In the case of DevOps the maturity model provided below is robust and can be used to improve capabilities broadly. The MDM (Master Data Model) framework is provided to simply demonstrate the wide range of domains that maturity models have been developed to cover.

### A. ITSM/ITIL Process Maturity Model

One of the most relevant maturity models from the perspective of an IT Operations team, for example, is the one that covers ITIL (IT Infrastructure Library). This model can also cover Service Management as shown in Figure 7 [Deloitte, 2004]. In assessing the process maturity of IT Operations organization this model can be beneficial. In particular, this model follows the same progression of earlier maturity models such as the SW-CMM yet it covers the IT Operations services and scope from Incident Management to Problem Management to Change Management which tend to exist outside of traditional CMM frameworks. The higher maturity levels of this model bring the user into more sophisticated applications of process including configuration management, SLA development, capacity management, and strategic partnerships. This framework provides a useful guide to the evolution of an IT Operations environment of process, tools, and staff capabilities. In addition, this model provides a clear progression through what can be a somewhat hard to follow set of best practices as provided by ITIL. Finally, this is also highly reusable in face of enhanced generations of ITIL. The current ITIL v3 is now giving way to ITIL 4 and this maturity model can be adapted to support this new approach with limited effort.

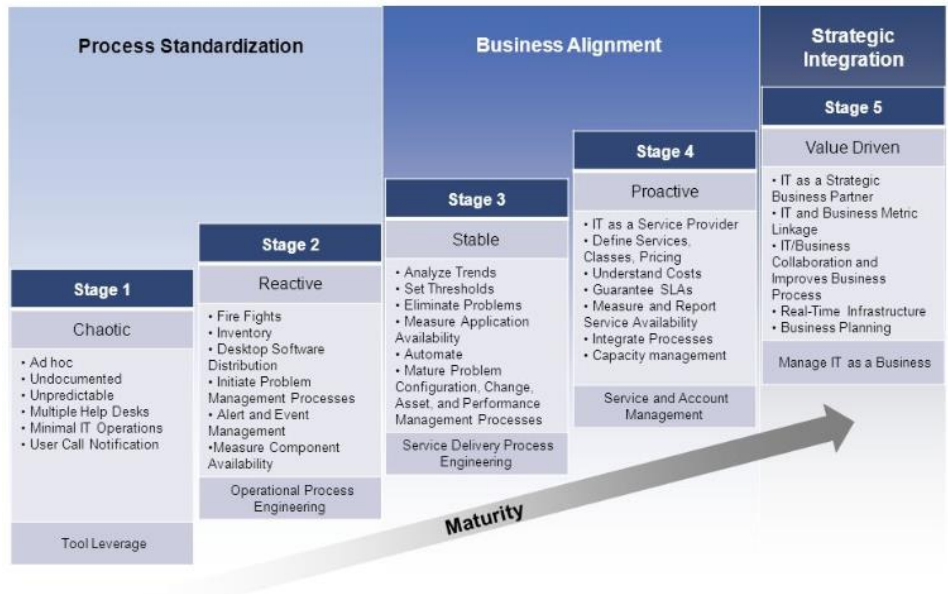

*Figure 7 – ITIL Maturity Model.*

B. *Agile Maturity Models*

For organizations that have responsibility for application development and may need to be assessed on Agile capabilities there are several extant models to choose from. Since about the year 2000 software development teams began practicing Agile to improve flexibility, customer/user satisfaction, and adaptability in the marketplace [Shore, 2007]. This mode of development is prevalent in many companies and may need to be accounted for when thinking about process improvement and planning for such improvements using a maturity framework.

The development process is important to discuss with respect to both Agile Development and DevOps. For Agile developers, instead of deploying software in large, scheduled releases, work is broken down into small, frequent iterations. Teams work to a deployable state, release as it's ready, and allow users to provide feedback on what works, what doesn't, and what can be improved. Scrum is a subset of Agile used mostly by development teams, which uses timed iterations on a product in two-week sprints. Custom Maturity Models for Scrum also exist such as the one from Yin [2011]. It is through DevOps methods which requires its own maturity view that these releases make it into progressive environments and are maintained in production operations.

Agile was influenced by many sources but key ideas came from the Toyota Production System developed by Taiichi Ohno [Womach, 1990]. This manufacturing process aimed to improve loss reduction and encourage sustainable production. It utilized visual signals to produce inventory exactly when it was needed (known as just-in-time production) and focused on optimizing the entire production system to minimize waste.

Converting this approach to Agile software development, the ideas of systems thinking and continuous improvement from Lean were threaded together into a fast-moving set of development practices to help organizations build healthy, innovative teams that sustainably deliver customer value. Implicit in this thinking is continuous improvement (known as kaizen in Japanese) and born of the PDCA or Shewhart cycles. Also implicit in Agile is that kaizen leads to higher maturity over time. As a result, many Agile maturity models have been developed [Humble, 2009]. Since there is no standard for such Agile maturity models this is a case where the user must assess the model which they will then use to assess their organization and process. A couple of examples here may help illustrate. First, a generalized maturity model for Agile is presented in Figure 8. Then a maturity model for use in assessing "build agility" can be applied as in Figure 9. We can see that this model follows a classical treatment of a Crosby/Humphrey Maturity Model being applied to Agile Development.

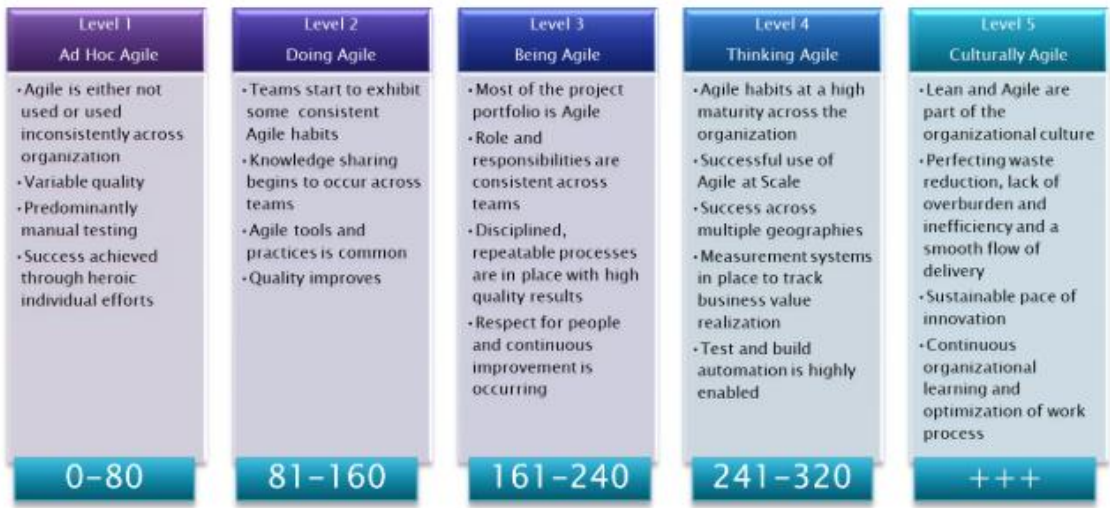

*Figure 8 –Cape Project Agile Maturity Model.*

| Practice | Build management and continuous integration | Environments and deployment | Release management and compliance | Testing | Data management |
|---|---|---|---|---|---|
| **Level 3 - Optimizing:** Focus on process improvement | Teams regularly meet to discuss integration problems and resolve them with automation, faster feedback, and better visibility. | All environments managed effectively. Provisioning fully automated. Virtualization used if applicable. | Operations and delivery teams regularly collaborate to manage risks and reduce cycle time. | Production rollbacks rare. Defects found and fixed immediately. | Release to release feedback loop of database performance and deployment process. |
| **Level 2 - Quantitatively managed:** Process measured and controlled | Build metrics gathered, made visible, and acted on. Builds are not left broken. | Orchestrated deployments managed. Release and rollback processes tested. | Environment and application health monitored and proactively managed. Cycle time monitored. | Quality metrics and trends tracked. Non functional requirements defined and measured. | Database upgrades and rollbacks tested with every deployment. Database performance monitored and optimized. |
| **Level 1 - Consistent:** Automated processes applied across whole application lifecycle | Automated build and test cycle every time a change is committed. Dependencies managed. Re-use of scripts and tools. | Fully automated, self-service push-button process for deploying software. Same process to deploy to every environment. | Change management and approvals processes defined and enforced. Regulatory and compliance conditions met. | Automated unit and acceptance tests, the latter written with testers. Testing part of development process. | Database changes performed automatically as part of deployment process. |
| **Level 0 – Repeatable:** Process documented and partly automated | Regular automated build and testing. Any build can be re-created from source control using automated process. | Automated deployment to some environments. Creation of new environments is cheap. All configuration externalised / versioned | Painful and infrequent, but reliable, releases. Limited traceability from requirements to release. | Automated tests written as part of story development. | Changes to databases done with automated scripts versioned with application. |
| **Level -1 – Regressive:** processes unrepeatable, poorly controlled, and reactive | Manual processes for building software. No management of artifacts and reports. | Manual process for deploying software. Environment-specific binaries. Environments provisioned manually. | Infrequent and unreliable releases. | Manual testing after development. | Data migrations unversioned and performed manually. |

*Figure 9 –ThoughtWorks Agile Build Maturity Model.*

C. *DevOps Maturity Models*

Similarly, DevOps has become a significant focus in the last several years. DevOps attempts to pick up where Agile leaves off in the development life-cycle. While Agile focused on efficiencies and lean methods for development this mostly stopped at the barrier between development and operations. DevOps by definition extends these same methods into IT Operations. This again prompts us to consider the need for a maturity model around DevOps to solve for the process engineering and standardization scope of the ITSM team. Effectively, when we combine Agile development and a DevOps approach to what was a traditional ITSM/ITIL environment a new framework may be called for. New integrated methods and tools are called for and progress in the implementation of these new models and solutions almost immediately generated new DevOps Maturity Models.

But prior to sharing those maturity models what in fact do we mean by DevOps? DevOps is the practice of operations and development engineers participating together in the entire service life-cycle, from design through the development process to production support. Here are a few basic principles of DevOps:

1. Adheres to Agile principles of Agile Manifesto [Beck, 2001].
2. Maintain best practices such as ITIL as they fit in an Agile mode.
3. Enable success through appropriate support tools and automation.

Naturally, to do this there will be many technical domains that come into play and this may also drive more than one maturity model. For the sake of brevity, we present only 2 models of the dozens available. The first is a generic maturity model for DevOps which ranges across many areas including culture, build/deploy release, and more (see Figure 10). Notice once again how this follows the classic Crosby/Humphrey approach. Indeed, there is nothing new under the sun. Additionally, in Figure 11 a focused Maturity Model detailing the progression of improvement applicable to the build and release domain is presented. Note that in both Agile and DevOps many of these specialty models can exist making usage even more complex.

*Figure 10 – A DevOps Maturity Model.*

*Figure 11 – A Specialized DevOps Build/Release Maturity Model from IBM.*

D. Master Data Management Maturity Model

A final maturity model of interest to demonstrate the wide range of use of the concept of maturity models is one specific to Master Data Management. This model can assist in guiding the evolution and operations of the process around improving the data model of data models, repositories, and master data stores. The model in Figure 12 shows promise in assisting in the creation of a strategic progress plan for MDMs and has been applied in that manner [Spruit, 2014].

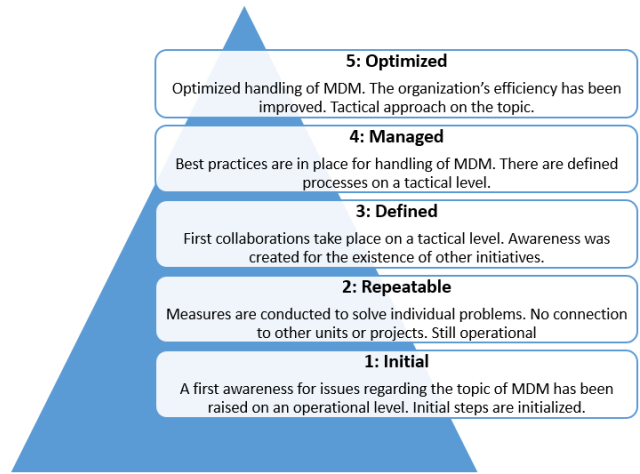

*Figure 12 - MDM Maturity Model (Adapted from Spruit, 2014).*

## V. APPLYING MATURITY MODELS FOR PROCESS IMPROVEMENT

The history of maturity models starting in the 1970s and maturing throughout the 1980s and 1990s has continued to be adapted to the latest methodologies bringing us to today's Agile and DevOps methodologies. This shows us that these models can be intuitive, comprehensive, extensible, and applicable to a wide range of process and organizational challenges. While some of the more robust frameworks can also be complex and to some may seem bureaucratic, they can also provide strong guidance from a best practices standpoint on where to put emphasis first regarding improvement plans. Maturity Models continue to be applied throughout the IT industry especially in core IT capabilities areas such as software, systems, service, and more. Additionally, maturity models for cybersecurity are popular and have been used effectively for several years.

While these models provide value in and of themselves to organize capability concepts and to order them in terms of improvement, the real value lies in their use in actually finding improvement areas through assessments and reviews and then planning and realizing those improvements. The literature on such improvements as guided by various maturity models is lengthy. Instead of attempting to recap this here a useful closing point will be illustrating where we began with Shewhart's SPC (Statistical Process Control) chart. In Figure 13 a sample SPC is provided (adapted from Card & Glass, 1990). This depicts the essence of process improvement – moving from one level of non-conformances to an improved level. At the most fundamental level this is what the maturity models are all attempting to do. By helping teams move up the maturity ladder these frameworks accomplish the reduction of non-conformances (or achieving an improvement in quality). Alternatively, we can view this as making incremental quality improvement steps as measured by an SPC and thereby reaching to the next rung on the maturity model which has been selected.

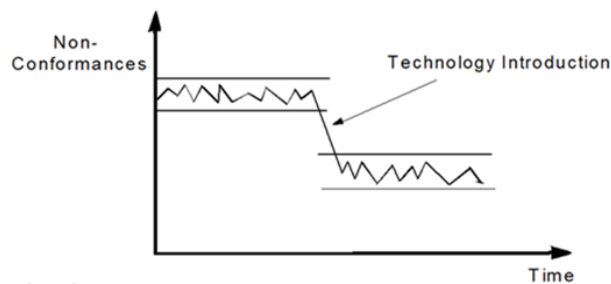

*Figure 13 - A Sample Shewhart Statistical Process Control Chart (Card, 1990).*

## VI. CRITICISMS OF MATURITY MODELS

This document would not provide an honest presentation of maturity models if a discussion of the critiques of these frameworks and assessments was not included. As it is there has been significant criticism of, for example, the CMM and CMMI frameworks. In fact, many parts of the software industry abandoned CMM/CMMI and adopted Agile methods precisely due to the taxing nature of the CMMI approach. This actually helped motivate the rise of the Agile Manifesto and the eventual demise of the complex plan-driven model of development for a more "lean" approach creating such methods as Scrum, Xtreme Programming, and DevOps. Today, raising the concept of a CMMI assessments to these teams is generally not welcome.

However, as has been shown above, the use of a custom Crosby/Humphries inspired maturity models tuned to Agile or DevOps are widely adopted if not standardized or supported with as much detailed implementation and assessment rigor.

## VII. CONCLUSION

This review began with a presentation of the development and spread of the fundamental process improvement methods used in manufacturing, engineering, and operations and eventually to IT. These methods were then shown to have a direct line to the development of business and quality maturity models. Finally, the formalization and dissemination of a diversity of maturity models was presented. From this review we can understand which types of models are best for application to such areas as Software Development, ITSM/ITIL, DevOps, or Project Management.

It is also clear from this review that there is no barrier to developing/customizing or applying a maturity model to nearly any business or technical domain for the purposes of understanding current state and developing a path to a future state. However, it is also clear that while CMMI is a well know model it is in fact not necessarily the right model to apply in its totality to each organization or domain. Instead, it is recommended to select appropriate frameworks as the case may call for it.

## VIII. ACKNOWLEDGEMENTS

My own journey with capability models and process improvement began at AT&T Bell Laboratories. There I was first exposed to these methods and was fortunate to lead a team of expert Process Engineers who taught me much in this domain. These people included Anil Midra, Bruce Pickens, Bruce Fetz, and Irwin Spock among others. I am also grateful to Columbia University for giving me the chance to teach Software Engineering to many fine students. This experience honed my knowledge of process and quality methods. Additionally, Mike Konrad of the SEI provided an opportunity for my team and I to help shape the first version of the CMMI which was a tremendous experience. Finally, in more recent roles I have also worked to deploy ITSM, Agile, and DevOps processes. This work has brought me into new and rewarding areas of applying maturity models.

## IX. REFERENCES

[1] ASQ, The History of Quality, https://asq.org/quality-resources/history-of-quality#20th-century, viewed 6/18/2019.
[2] Bacon, Francis, **The Essays of Francis Bacon**, CreateSpace Independent Publishing Platform, January 19, 2013, originally published 1696.
[3] Beck, K., et. al., "*Manifesto for Agile Software Development*", https://agilemanifesto.org/, 2001.
[4] Card, R., Glass, D. **Measuring Software Design Quality**, Prentice Hall, NJ, 1990.
[5] Chrissis, M. B., et. al., **CMMI Guidelines for Process Integration and Product Improvement**, 2nd Edition, Addison-Wesley, Upper Saddle River, NJ, 2007.
[6] Chrissis, Mary Beth, et., al., **CMMI for Development: Guidelines for Process Integration and Product Improvement**, 3rd Edition, SEI Series in Software Engineering, Addison-Wesley Professional, Mar 20, 2011.
[7] Chrissis, Mary Beth, et., al., **CMMI: Guidelines for Process Integration and Product Improvement**, Addison-Wesley Professional, March 6, 2003.
[8] CMMI Institute, "*CMMI Institute Publishes A Guide to Scrum and CMMI: Improving Agile Performance with CMMI*", Press Releases, 2017/01/04.
[9] Crosby, Philip B., **Quality is Free: The Art of Making Quality Certain**, McGraw-Hill Book Company, New York, 1979.
[10] Deloitte Consulting, "*Commonwealth of Massachusetts ITC Consolidation Initiative*", September 3, 2009.
[11] Deming, W. Edwards, **Out of the Crisis**, The MIT Press, Reprint Edition, 2000, Originally published 1982.
[12] Galileo, Galilei, **Two New Sciences/A History of Free Fall**, Combined Edition, Stillman Drake (Author, Translator), Wall & Emerson, Inc., September 1, 2000, originally published 1638.
[13] Gresse von Wangenheim, Christiane, "*Creating Software Process Capability/Maturity Models*", **IEEE Software**, 27(4), August 2010.
[14] Humble, J., et. al., "*The Agile Maturity Model: Applied to Building and Releasing Software*", **ThoughtWorks Studios**, September 2009, https://info.thoughtworks.com/rs/thoughtworks2/images/agile_maturity_model.pdf.
[15] Ishikawa, Kaoru, **Guide to Quality Control**, Asian Productivity Organization, Tokyo, 1968.
[16] Ishikawa, Kaoru, **What is Total Quality Control? The Japanese Way,** [Originally titled: TQC towa Nanika—Nipponteki Hinshitsu Kanri], D. J. Lu (trans.), New Jersey: Prentice Hall, 1985, Originally published 1981.
[17] Juran, Joseph, **Quality Control Handbook**, New York, New York: McGraw-Hill, 1951.
[18] Moen, Ronald, "*PDSA History*", **16th Deming Research Seminar**, Feb 2010.
[19] Nolan, Richard, "*Managing the Crisis in Data Processing*", **HBR**, 1973
[20] Paulk, M. C., et. al., **The Capability Maturity Model: Guidelines for Improving the Software Process**, Addison-Wesley, Reading, Massachusetts, 1994.
[21] Paulk, M., "*A History of the Capability Maturity Model for Software*", Software Engineering Institute, 2001.
[22] Radice, R., et. al., "*A Programming Process Study*", **IBM Systems Journal**, Vol. 24, No. 2, 1985.

[23] Shewhart, Walter, **Economic Control of Quality Of Manufactured Product,** Martino Fine Books, April 25, 2015, Originally published 1931.
[24] Shore, James, & Warden, Shane, **The Art of Agile Development: Pragmatic Guide to Agile Software Development,** O'Reilly Media; 1 edition, Nov 5, 2007.
[25] Spruit, Marco, "*MD3M: The master data management maturity model*", **Computers in Human Behavior**, October 2014, DOI: 10.1016/j.chb.2014.09.030
[26] Taylor, Frederick Winslow, **The Principles of Scientific Management**, CreateSpace Independent Publishing Platform, October 27, 2010, originally published 1911.
[27] Womach, J., et., al., **The Machine That Changed the World**, Free Press, New York, 1990.
[28] Yin, Alexandre, et. al., *Scrum Maturity Model*, ReseachGate.org, January 2011

X. ABOUT THE AUTHOR

James Cusick is an IT leader with over 30 years of experience in Software Engineering, Information Security, IT Operations, Process Engineering, Project Management, and Applied Research. He is currently Senior Director IT Strategy and Operations & Distinguished Engineer with a global information services firm. In earlier roles with the company James led the global IT Service Management Process Engineering team, acted as Busines Unit CISO, and led IT Operations. Previously, James held leadership and technical roles with Dell Services, Lucent Bell Laboratories, and AT&T Laboratories.

James is currently a Board Trustee at the Henry George School of Social Science where he researches topics at the intersection of Innovation and Economics. James was also an Adjunct Assistant Professor of Computer Science at Columbia University where he taught Software Engineering. James has also published over 100 papers and talks in his fields of interest including two recent books on IT and Software Engineering. James is a graduate of the University of California at Santa Barbara and Columbia University in New York City, an IEEE Computer Society Member, and a certified PMP (Project Management Professional). Contact James at j.cusick@computer.org.